\documentclass{article}
\usepackage{times,amsfonts,cite}

\begin{document}

\title{A selfadjoint variant of the time operator}

\author{Rafael de la Madrid \\ [0.1ex]
\small{{\it Department of Physics, University of California at San Diego, La 
Jolla, CA 92093}} \\ [-0.5ex]
\small{E-mail: \texttt{rafa@physics.ucsd.edu}} \\ [2ex]
Jos\'e M.~Isidro \\ [0.1ex]
\small{{\it Instituto de F\'{\i}sica Corpuscular (CSIC--UVEG)
Apartado de Correos 22085, Valencia 46071, Spain}} \\ [-0.5ex]
\small{E-mail: \texttt{jmisidro@ific.uv.es}}}

\maketitle

\begin{abstract}
\noindent We study the selfadjoint time operator recently constructed by one 
of the authors. We will show that this time operator must be interpreted as 
a ``selfadjoint variant'' of the time operator.
\end{abstract}

\section{Introduction}
\label{sec:Intro}

Ever since Pauli~\cite{PAULI} proved that there cannot exist a selfadjoint 
time operator $T$ that commutes canonically with a bounded-from-below 
Hamiltonian $H$, there have been many attempts to construct meaningful time 
operators in quantum mechanics.

One way to circumvent Pauli's objection is to work directly with the 
non-selfadjoint $T$ by way of Positive Operator Valued Measures (POVMs), 
see e.g.~\cite{BUSCH,GIANNITRAPANI,EGUSQUIZA,JULVE} and
especially the reviews in~\cite{TIME}. Another way is to manipulate the 
original time operator $T$ until we obtain a meaningful, selfadjoint time
operator. The resulting operators are called ``selfadjoint variants'' of 
the time operator~\cite{LEAVENS}. In this second category, we find the 
``selfadjoint variants'' of the time-of-arrival operator of 
Razavi~\cite{RAZAVI}, that of Grot {\it et al.}~\cite{ROVELLI} (see 
also~\cite{PAUL}), that
of Kijowski~\cite{KIJOWSKI,DELGADO}, and that of Galapon 
{\it et al.}~\cite{GALAPON}. Although these operators are all selfadjoint, 
they lack other desirable properties such as covariance~\cite{LEAVENS} or
a canonical commutation relation with $H$. There are time
operators, such as the dwell time~\cite{DAMBO} or the time delay~\cite{AMREIN},
that are selfadjoint and commute with the Hamiltonian, and therefore Pauli's 
theorem does not apply to them. In the relativistic domain, it has been 
recently reported that the time-of-arrival operator has self-adjoint 
extensions~\cite{WANG}.

The efforts to construct time operators parallel those to construct 
phase operators. It is known that there cannot exist a selfadjoint phase
operator that canonically commutes with the number 
operator~\cite{SUSSKIND}. Similarly to the time operator, one has to either 
work with the original phase operator using POVMs~\cite{HALL} or 
construct selfadjoint variants of it, see e.g.~\cite{NIETO,KASTRUP}.

In a recent paper~\cite{PLA}, one of us has proposed a selfadjoint time 
operator, denoted $T_{\sqrt{}}$. We wish here to find the place of
$T_{\sqrt{}}$ among the other time observables. Its properties, 
especially the fact that $T_{\sqrt{}}$ does not canonically commute with
the Hamiltonian, will lead us to conclude that $T_{\sqrt{}}$ is 
another ``selfadjoint variant'' of the time operator. We will also see that
incorporating $T_{\sqrt{}}$ into the algebra of observables leads to a
variant of the Heisenberg algebra.

\section{Construction of $T_{\sqrt{}}$}
\label{qmhft}

Let us first summarize the construction of $T_{\sqrt{}}$ 
following~\cite{PLA}. We use the holomorphic Fourier transformation (HFT)
\begin{eqnarray}
      \varphi(t)&=&{1\over \sqrt{2\pi\hbar}}\int_0^{\infty}{\rm d} E\, 
             f(E)\,{\rm e} ^{{{\rm i}\over\hbar}Et}  \cr  \cr 
      f(E)&=&{1\over \sqrt{2\pi\hbar}}\int_{-\infty}^{\infty}{\rm d} t\, 
        \varphi(t)\, {\rm e}^{-{ {\rm i}\over \hbar}Et}  
             \label{qreal}
\end{eqnarray}
where $t$ is a complex variable defined on the upper half plane $\mathbb{H}$,
and $E$ is a real variable defined on the positive axis $[0,\infty)$. We can
promote $t$ and $E$ to quantum operators $T$ and $H$ by defining
\begin{equation}
    (Hf)(E):=E\,f(E)\, , \qquad 
     (Tf)(E):={\rm i}\hbar\,{{\rm d} f\over {\rm d} E} \, .
      \label{pdef}
\end{equation}
The HFT~(\ref{qreal}) provides a conjugate representation of 
Eq.~(\ref{pdef}):
\begin{equation}
     (H\varphi)(t)=-{\rm i}\hbar\, {{\rm d}\varphi\over {\rm d} t} \, ,\qquad 
      (T\varphi)(t)=t\,\varphi(t) \, .
            \label{zdef}
\end{equation}
Obviously, $T$ and $H$ satisfy Heisenberg's commutation relation: 
\begin{equation}
    [T, H]= {\rm i}\hbar\,{\bf 1} \, .
       \label{hei}
\end{equation}
Equation~(\ref{hei}) is valid in the intersection of the domains of $T$ and
$H$, which we are going to specify now.

The domain of $H$ is
\begin{equation} 
           D(H)=\{f\in L^2([0,\infty),{\rm d}E):\; 
                 Ef\in L^2([0,\infty),{\rm d}E)  \} \, ,
          \label{domq}
\end{equation}
which is dense in $L^2([0,\infty),{\rm d}E)$. On $D(H)$, the operator $H$ is 
symmetric,
\begin{equation}
          \langle g\vert H f\rangle= \langle Hg\vert f\rangle \, .
          \label{barp}
\end{equation} 
The defect indices $d_{\pm}$ of $H$ are equal to zero, and therefore $H$
is selfadjoint~\cite{YOSIDA}. Its spectrum is the positive real line,
\begin{equation}
          \sigma (H)=[0,\infty) \, .
       \label{pspectra}
\end{equation}

The properties of $T$ are subtler. A straightforward calculation yields
\begin{equation} 
       \langle g\vert Tf\rangle={\rm i}\hbar\, f(0)g^*(0) + \langle Tg\vert f\rangle \, ,
        \label{barzop}
\end{equation} 
so $T$ is symmetric on the domain 
\begin{equation}
       D(T)=\{ f\in L^2([0,\infty),{\rm d}E):  f\in AC([0,\infty )) \, , \ 
             f' \in L^2([0,\infty),{\rm d}E) \, , \  f(0)=0 \} \, ,
         \label{domz}
\end{equation}
where $AC([0,\infty ))$ denotes the space of absolutely continuous functions
on the positive real line. The adjoint $T^{\dagger}$ also acts as 
${\rm i}\hbar\,{\rm d}/{\rm d} t$, but on the following domain:
\begin{equation} 
            D(T^{\dagger})=\{ f\in L^2([0,\infty),{\rm d}E): 
                            f\in AC([0,\infty )) \, \  
                             f' \in L^2([0,\infty),{\rm d}E) \}  \, ,
              \label{domzdag}
\end{equation}
where the condition $f(0)=0$ has been lifted. The defect indices of $T$ are
$d_{+}(T)=0$, $d_{-}(T)=1$. Because a symmetric operator has selfadjoint 
extensions if and only if its defect indices are equal, we conclude 
that $T$ admits no selfadjoint extension. Since $T$ is not selfadjoint,
the spectrum of $T$ cannot be real and in fact is
\begin{equation}
         \sigma (T)=\mathbb{H}\cup \mathbb{R} \, .
              \label{zspectra}
\end{equation}

Even though $T$ is not selfadjoint and its spectrum includes complex numbers,
a way was found in~\cite{PLA} to construct a selfadjoint time operator with
real spectrum out of $T$. In order to do so, we need to construct first 
the operator $T^2$. This operator acts as
\begin{equation}
       T^2=-\hbar ^2 \frac{{\rm d}^2}{{\rm d}E^2} \, .
          \label{tequs}
\end{equation}
The operator $T^2$ is symmetric, and its defect indices are
$d_-(T^2)=d_+(T^2)=1$. Hence, $T^2$ has infinitely many selfadjoint
extensions. The selfadjoint extension used in~\cite{PLA} has the following
domain:
\begin{equation}
   {D}(T_F^2) = \{ f\in L^2([0,\infty),{\rm d}E) : f(0)=0 \, , \ 
                      f\in AC^2([0,\infty))\, , \
                      f''\in L^2([0,\infty),{\rm d}E) \} \, ,
      \label{domainF}
\end{equation}
where $AC^2([0,\infty))$ stands for the space of functions whose first 
derivative is absolutely continuous. The operator~(\ref{tequs})
acting on the domain~(\ref{domainF}) is simply the Friedrichs 
extension~\cite{YOSIDA}, and its spectrum coincides with the positive real 
line:
\begin{equation}
       \sigma (T^2_F)=[0,\infty) \, .
         \label{ptwospectra}
\end{equation}

The crucial point is that the square root of the Friedrichs extension allows 
us to define a selfadjoint time operator:
\begin{equation}  
         T_{\sqrt{}}:=+\sqrt{T^2_F} \, .
              \label{square}
\end{equation}
Because it is the square root of a selfadjoint operator, $T_{\sqrt{}}$ is 
selfadjoint. In particular, its spectrum is real and coincides with the 
positive real line:
\begin{equation}
        \sigma (T_{\sqrt{}})=[0,\infty) \, .
             \label{spectrappm}
\end{equation}

\section{Properties of $T_{\sqrt{}}$}
\label{sec:proof}

The properties of $T_{\sqrt{}}$ are determined by those of
$T_F^2$ through the spectral theorems. Thus, in order to
obtain the properties of $T_{\sqrt{}}$, we need first to obtain the properties
of $T_F^2$. From Eq.~(\ref{square}), one may naively think 
that $T_{\sqrt{}}$ is not only selfadjoint but also satisfies all the nice 
properties of $T$ such as Eq.~(\ref{hei}). On the contrary, we will see that 
the properties of $T_{\sqrt{}}$ differ drastically from those of $T$, the
reason being that the operation of taking the Friedrichs extension does not 
commute with the operation of taking the square root.

In the energy representation, the eigenfunctions of $T_F^2$ are given by 
\begin{equation}
        \langle E|t\rangle = \sqrt{\frac{2}{\pi \hbar}} \sin (Et/\hbar) \, .
               \label{sine}
\end{equation}
These eigenfunctions are delta-normalized,
\begin{equation}
        \int_0^{\infty}{\rm d}E \,  \langle t|E\rangle \langle E|t'\rangle 
          = \delta (t-t') \, ,
    \label{deltanorm}
\end{equation}
and their corresponding eigenvalue is $t^2$,
\begin{equation}
        T_F^2|t\rangle = t^2|t\rangle  \, .
\end{equation}
Thus, the spectral representation of $T_F^2$ is
\begin{equation}
        T_F^2=\int_0^{\infty}{\rm d}t \, t^2  |t\rangle \langle t|  \, .
\end{equation}

By contrast to the original time operator $T={\rm i}\hbar {\rm d}/{\rm d}E$, 
the eigenfunctions of $T_F^2$ can be delta-normalized, see 
Eq.~(\ref{deltanorm}), and therefore one can construct a time representation 
associated with $T_F^2$. The unitary operator $U$ that 
transforms from the energy representation into the time representation reads
\begin{equation}
      (Uf)(t)=\int_0^{\infty}{\rm d}E \, \sqrt{\frac{2}{\pi \hbar}}
                     \sin (Et/\hbar) f(E) \, .
            \label{Udoepr}
\end{equation}
The operator $U$ brings the original Hilbert space 
$L^2([0,\infty ),{\rm d}E)$ onto the Hilbert space of the time 
representation $L^2([0,\infty ),{\rm d}t)$. In 
the time representation, $T_F^2$ acts simply as multiplication by $t^2$, 
whereas the squared Hamiltonian acts as 
\begin{equation}
        H^2 = -\hbar ^2 \frac{{\rm d}^2}{{\rm d}t^2} \, ,
\end{equation}
as can be easily seen by using Eq.~(\ref{Udoepr}). Note that the Hamiltonian 
$H$ does not have a simple form in the time representation. In particular, 
$H$ does not act as ${\rm i}\hbar {\rm d}/{\rm d}t$.

The time operator $T_{\sqrt{}}$ is unambiguously defined by the spectral
theorems. It has the same eigenfunctions as $T_F^2$, but the corresponding
eigenvalue is $t$:
\begin{equation}
        T_{\sqrt{}}\,|t\rangle = t |t\rangle  \, ,
\end{equation}
that is, its spectral representation is
\begin{equation}
        T_{\sqrt{}}=\int_0^{\infty}{\rm d}t \, t |t\rangle \langle t|  \, .
\end{equation}
Note that, in particular, the eigenfunctions of $T_{\sqrt{}}$ are the 
sine functions~(\ref{sine}) rather than the original eigenfunctions
${\rm e}^{{\rm i}Et/\hbar}$ of the operator $T$, and that $T_{\sqrt{}}$ 
acts as multiplication by $t$ in the time representation. Note also that
the time representation associated with $T_{\sqrt{}}$ is a well-defined 
representation, not just a POVM as is the case of the original time 
operator $T$.
        
Although selfadjoint, $T_{\sqrt{}}$ does not canonically commute with the
Hamiltonian, because $T_{\sqrt{}}$ does not act as 
${\rm i}\hbar {\rm d}/{\rm d}E$ in the energy representation (or, what is the 
same, $H$ does not act as ${\rm i}\hbar {\rm d}/{\rm d}t$ in the 
time representation). Thus, even though we generated a selfadjoint operator
from our original time operator $T$, the resulting $T_{\sqrt{}}$ does not
satisfy Eq.~(\ref{hei}). Hence $T_{\sqrt{}}$ is a ``selfadjoint variant'' 
of the time operator.

Since it is selfadjoint, $T_{\sqrt{}}$ is an observable quantity according
to the standard rules of quantum mechanics. We can therefore include it in the
algebra of observables. In this paper, the other observable we are considering 
is the Hamiltonian, and therefore the simplest algebra one can think of is 
that generated by $T_{\sqrt{}}$, $H$ and the identity 1. However, because 
$T_{\sqrt{}}$ and $H$ do not commute canonically, the commutator
\begin{equation}
      I=[T_{\sqrt{}},H]
\end{equation}
is not given in terms of the generators $T_{\sqrt{}}$, $H$ and 1. In order
to close the algebra, we must therefore include $I$ as one of the 
generators:
\begin{equation}
          {\cal A}\equiv \{  T_{\sqrt{}}, \, H, I, \, 1 \} \, .
        \label{alaa}
\end{equation}
If we had a Heisenberg algebra, then $I$ would equal ${\rm i}\hbar 1$, and it 
is in this sense
that the algebra $\cal A$ is a variant of the Heisenberg algebra. Note however 
that although the Jacobi identity
\begin{equation}
       [[A,B],C] + [[B,C],A]+ [[C,A],B] = 0
\end{equation}
is satisfied when $A$, $B$ and $C$ belong to $\cal A$, the commutators
of $I$ with $H$ and $T_{\sqrt{}}$ cannot be written in terms of linear 
combinations of the generators of $\cal A$. Hence, $\cal A$ is an enveloping
algebra rather than a Lie algebra.

In quantum mechanics, we almost always use Lie algebras. If we
insisted that the algebra of observables be a Lie algebra, we should
modify~(\ref{alaa}) accordingly. The simplest Lie algebra that includes
$T_{\sqrt{}}$ is
\begin{equation}  
   {\cal A}' =\{ T_{\sqrt{}}^2, \, T, \, H, \, 1 \}  \, .
\end{equation}
One can check that ${\cal A}'$ is indeed a Lie algebra. The price to pay,
however, is that we are forced to include the non-selfadjoint operator 
$T$ and that we cannot include $T_{\sqrt{}}$ but $T_{\sqrt{}}^2$.

Thus, we can either include $T_{\sqrt{}}$ in a non-enveloping 
algebra or we can include $T_{\sqrt{}}^2$ and the non-selfadjoint $T$ in a 
Lie algebra. Further progress must be made to see if the limitations of these 
algebras can be somehow overcome.

\section{Conclusion}
\label{sec:conc}

In order to summarize the results of this paper, we compare the
properties of $T$ with those of $T_{\sqrt{}}$. Whereas the operator $T$ is not 
selfadjoint and its spectrum is the closed, upper half of the complex plane,
the operator $T_{\sqrt{}}$ is selfadjoint and its spectrum is the positive 
real line. There is a time representation associated with the operator 
$T_{\sqrt{}}$ on which $T_{\sqrt{}}$ acts as multiplication by $t$, whereas
the POVM associated with $T$ does not provide a well-defined time 
representation. Whereas $T$ commutes canonically with $H$, $T_{\sqrt{}}$ does
not. 

%The operators $T$, $H$ and ${\bf 1}$ furnish a representation of the
%Heisenberg algebra, whereas $T_{\sqrt{}}$, $H$ and ${\bf 1}$ furnish a 
%variant thereof.

Finally, we note that by applying the HFT to the radial momentum operator 
in three dimensions and to the momentum operator in the half-line, we can
also construct selfadjoint variants of these operators.

\vskip1cm
\noindent
{\bf Acknowledgments}. The authors wish to thank Gonzalo Muga and 
I\~nigo Egusquiza for enlightening discussions. The work of J.M.I.~has been 
supported by Ministerio de Educaci\'{o}n y Ciencia (Spain) through grant 
FIS2005--02761, by Generalitat Valenciana, by EU FEDER funds and by EU network 
MRTN--CT--2004--005104 ({\it Constituents, Fundamental Forces and Symmetries 
of the Universe}). R.M.~acknowledges the financial support of MEC and DOE.

\end{document}